\documentclass[letter]{aa}
\usepackage[varg]{txfonts}

\usepackage[unicode=true,psdextra]{hyperref}
\usepackage{natbib}
\usepackage{graphicx}

\begin{document}

\title{Hidden by a star: the redshift and the offset broad line of the Flat Spectrum Radio Quasar PKS~0903-57 \thanks{Based on observations collected at the European Organization for Astronomical Research in the Southern Hemisphere, Chile, under programs P10822CJ.002 and P112.25SN.001. The raw FITS data files are available in the ESO archive. Based on observations made with the Southern African Large Telescope~(SALT) under program 2020-1-SCI-027 (PI E.~Kasai).} }

\author{
    P. Goldoni  \inst{1} 
     \and
    C. Boisson \inst{2}
     \and
    S. Pita \inst{3}
     \and
    F. D'Ammando \inst{4}
     \and
    E. Kasai \inst{5} 
     \and
    W. Max-Moerbeck \inst{6}
     \and
    M. Backes \inst{5,7}
     \and
     G. Cotter \inst{8}
    }

        \institute{
Universit\'{e} Paris Cit\'{e}, CNRS, CEA, Astroparticule et Cosmologie, F-75013 Paris, France  
 \and
Laboratoire Univers et Th\'{e}ories, Observatoire de Paris, Universit\'{e} PSL, Universit\'{e} Paris Cit\'{e}, CNRS, F-92190 Meudon, France
 \and
Universit\'{e} Paris Cit\'{e}, CNRS, Astroparticule et Cosmologie, F-75013 Paris, France
 \and
INAF - Istituto di Radioastronomia, Via Gobetti 101, I-40129 Bologna, Italy 
 \and
Department of Physics, Chemistry \& Material Science, University of Namibia, Private Bag 13301, Windhoek, Namibia
 \and
Departamento de Astronom\'{i}a, Universidad de Chile, Camino El Observatorio 1515, Las Condes, Santiago, Chile
 \and
Centre for Space Research, North-West University, Private Bag X6001, Potchefstroom 2520, South Africa
 \and
Department of Physics, University of Oxford, Keble Road, Oxford OX1 3RH, United Kingdom
 }

\offprints{P. Goldoni \email{goldoni@apc.in2p3.fr}}

\date{Received/Accepted}

\abstract
{PKS~0903-57 is a little-studied $\gamma$-ray blazar which has recently attracted considerable interest due to the strong flaring episodes observed since 2020 in HE (100~MeV $\le E \le 100$~GeV) and VHE (100~GeV $\le E \le 10$~TeV) $\gamma$-rays. Its nature and properties are still not well determined. In particular, it is unclear whether PKS~0903-57 is a BL~Lac or a Flat Spectrum Radio Quasar (FSRQ), while its redshift estimation relies on a possibly misassociated low signal-to-noise spectrum.}
{We aim to reliably measure the redshift of the blazar and to determine its spectral type and luminosity in the optical range.}
{We performed spectroscopy of the optical counterpart of the blazar using the South African Large Telescope (SALT) and the Very Large Telescope (VLT) and monitored it photometrically with the Rapid Eye Mount (REM) telescope.}
{We firmly measured the redshift of the blazar as $z= 0.2621\pm0.0006$ thanks to the detection of five narrow optical lines. The detection of a symmetric broad H$\alpha$ line with Full Width at Half Maximum (FWHM) of $4020 \pm 30$~km/s together with a jet-dominated continuum leads us to classify it as a FSRQ. Finally, we detected with high significance a redshift offset ($\sim 1500$~km/s) between the broad line and the host. This is the first time that such an offset is unequivocally detected in a VHE blazar, possibly pointing to a very peculiar accretion configuration, a merging system, or a recoiling Black Hole.}
{}


\maketitle

\nolinenumbers

\section{Introduction}
  Blazars are the brightest persistent $\gamma$-ray sources in the HE and VHE sky, their redshift is often difficult to measure due, in the case of BL~Lacs, to their featureless continuum-dominated spectrum.  In the perspective of the future operations of the Cherenkov Telescope Array Observatory (CTAO),
  we are performing a multi-year observation campaign aimed at measuring or constraining the redshift of blazars likely to be detected with CTAO \citep{Gol21,Kas23b,Dam24}. As of today, about half of the {\it Fermi}-LAT blazars lack redshifts or only have old, low S/N spectra which do not always yield reliable redshifts or spectral classification.
  
   PKS~0903-57 is a {\it Fermi}-LAT detected blazar with a rather hard (spectral index $\sim 2.2$) spectrum which is one of the targets of our campaign. It displayed $\gamma$-ray bright flares in 2020 through 2022 \citep{Bus20,Wag20,Lam22} reaching brightness levels of 4--7 Crab in the VHE domain, which prompted multiple multiwavelength publications \citep{Mon21,Shah21}.  Despite this, in the 4LAC catalog \citep{Aje22} it is still classified as a Blazar with Uncertain Classification (BCU). PKS~0903-57 was discovered in radio by \citet{Bol64}, its radio position is $RA=09$h04m$53.17$s$\pm0$\farcs$1$, $DEC=-57^{\circ}35^{'}05$\farcs$8\pm0$\farcs$15$ \citep{Cha20}. The source is located near the Galactic Plane ($l\sim276^{\circ}$, $b\sim-7^{\circ}$), with strong line-of-sight absorption, $E(B-V)=0.282$ \citep{Sch11}. 
The optical counterpart has a magnitude $B=17.6$ \citep{Whi87}. The only redshift measurement was reported by \citet{Tho90}. Their spectrum has a low signal-to-noise with a power law shape displaying two weak features at 4751 and 6306~\AA~identified as MgII and [OII] at redshift $z=0.695$. This is the usually accepted value, nevertheless, there appears to be some ambiguity about that identification. The source with that redshift is indeed 4\arcsec off the radio counterpart. 
  ATCA radio observations detected a one-sided arcsec-scale jet to the northeast of the core while {\it Chandra} X-ray observations detected the core and weak emission associated with the radio jet \citep{Mar05}. Although the X-ray position is not reported in that work, examination of the contour plot of the radio and X-ray detection (see their Figure~1,f) points toward the optical counterpart reported by \citet{Whi87} as the correct one. The {\it Gaia} DR3 catalog \citep{Gai23} reports near the radio position two sources separated by 0\farcs67. The fainter one has magnitude $G=17.87\pm0.03$ and is at the radio position, the brighter one has magnitude $G=15.94\pm0.01$. The results just described suggest that the faint {\it Gaia} source is the blazar counterpart while the bright one is a Galactic star along our line of sight. If this hypothesis is correct, the spectra of the two objects are very different with absorption lines from the star and a power law spectrum with possibly redshifted emission lines from the blazar. Therefore, despite the very small separation, depending on the relative brightness of the two sources, it should be possible to detect the signatures of both of them. We report on the observations we performed in order to test this hypothesis. 
 
   For all calculations, we used a cosmology with $\Omega_M = 0.27$, $\Omega_{\Lambda} = 0.73$, and $H_0 = 70$~km~s$^{-1}$~Mpc$^{-1}$. All wavelengths are in air. All magnitudes are in the Vega system.
  
\section{Observations and Results}

\begin{table*}[!]
\small
\caption{\label{tabobs} List of parameters of the spectroscopic observations for PKS~0903-57.} 
\centering
\begin{tabular}{lccccccccc}
\hline\hline
Telescope &    Instrument &   Range        & Resolution &Slit          & Pos.Ang.  & Start Time & Exp.  & Airm. & Seeing      \\
         &                &    (\AA)         &            &(\arcsec).    & (degrees)  &  (UTC)          &  (s) &       &   (\arcsec)      \\  
\hline\hline
SALT       & RSS/PG0900    &  4500--7500  & 1000    &  2              & 0           & 2020-05-14T19:08:33   & 2080 &   1.34  &   1.80   \\      
SALT       & RSS/PG0900   &  4500--7500   & 1000    &2             & 90          & 2020-05-22T18:25:33   & 2100 &   1.27  &  1.50    \\           
VLT         &  FORS/600RI    &  5200--8300  & 1000    &0.7          & 18     & 2022-03-25T02:32:23   & 2680 &    1.24  &   0.70   \\           
VLT         &  FORS/600RI    &  5200--8300  & 1000    &0.7          & 108   & 2022-03-26T00:30:54   & 2680 &    1.21  &   0.68    \\           
VLT         &  FORS/600RI    &  5250--8400  & 1000    &0.5          & 108   & 2024-01-07T06:58:15   & 2680 &    1.20  &   0.56    \\           
VLT         &  FORS/1028z    &  7800--9000  & 2500    & 0.5          & 108   & 2024-01-07T07:52:07   & 2680 &    1.24  &   0.67     \\           
\hline\hline
    
\end{tabular}
 \tablefoot{Position angle = Slit position angle (degrees) (North=0 East=90).}
\end{table*}

  After the detection of the 2020 flare of PKS~0903-57, we triggered observations with the SALT telescope using the Robert Stobie Spectrograph (RSS) \citep{Bur03} to catch the blazar in a bright state. Our observations were centered at the radio position with different slit position angles defined as 0~degrees when the slit is placed in the N-S direction  and 90~degrees when in the E-W direction (see \autoref{tabobs} and \autoref{figslits}).
 
 The SALT spectra obtained from the two different slit orientations did not show any noticeable difference and presented only features consistent with a star of type F or G  (see \autoref{figSALT}).
 In particular, we could not detect the putative emission features at 4751 and 6306~\AA~with limits on the equivalent width (EW) of $\sim 0.6~\AA$. We conclude that at this epoch and in this configuration the blazar optical flux was too weak to be detectable over the flux of the star.

   To complement these spectra we found in the European Southern Observatory (ESO) archive observations of our target\footnote{ProgID 0100.A-0588(A), PI V. Moss} taken with the ESO Faint Object Spectrograph and Camera EFOSC2 \citep{Buz84} on March 12, 2018, using grism Gr~13 ($\Delta \lambda/\lambda \sim 300$; range 3700--9300~\AA). These spectra have lower S/N but a wider wavelength range and still do not show any Active Galactic Nucleus (AGN) signature or emission lines of any kind. Combining the two spectra, we compared them to main sequence star templates from the X-shooter Spectral Library DR3 \citep{Ver22} finding that the star is likely of spectral type F6. By reddening and scaling the template using the {\it Gaia} magnitude ($G=15.94\pm0.01$), we obtained a photometric template of the star. Under the assumption that the star is not variable, which is reasonable considering its ordinary spectrum, this template can be subtracted from unresolved photometry of the two objects to obtain the blazar flux.

 We performed further observations in 2022 at ESO using the FOcal Reducer and low dispersion Spectrograph (FORS2) \citep{App98}. Given the SALT results, we modified our configurations using a 0\farcs7 slit, asking for good (0\farcs8) seeing. The pointing was achieved using blind offsets from a nearby star. We used position angles $PA=18^{\circ}$ and $PA=108^{\circ}$ respectively parallel and perpendicular to the line joining the two objects. The observations were split into six observing blocks each about 890 seconds, three for each position angle.
The spectra were still dominated by the bright star \citep{Kas23a}. The spectra at $PA=18^{\circ}$ did not yield any new feature, however, the spectra at $PA=108^{\circ}$ displayed a weak, narrow emission line at $\sim6318~\AA$~with EW$=0.7\pm0.2$~\AA. It can be interpreted as [OII] or [OIII]b \citep[see e.g.][]{Kas23b}, the first hypothesis would yield $z \sim 0.695$, i.e. the redshift proposed by \citet{Tho90}, the second $z\sim0.262$. Among the spectra at $PA=108^{\circ}$, the line was stronger in the first two observing blocks where the seeing was 0\farcs65 and was weaker in the third one (seeing 0\farcs76) suggesting that very good seeing was a key factor for its detection \citep[see][Figure~2 on the right]{Kas23a}. We thus asked for another observation to be performed with the slit at $PA=108^{\circ}$ using a 0\farcs5 slit in exceptional (0\farcs5) seeing and also for three observing blocks using the Grism Gr1028z with coverage down to 9000~\AA~to check for the presence of other lines.

\begin{figure}[!]
   \centering
 \includegraphics[scale=0.32]{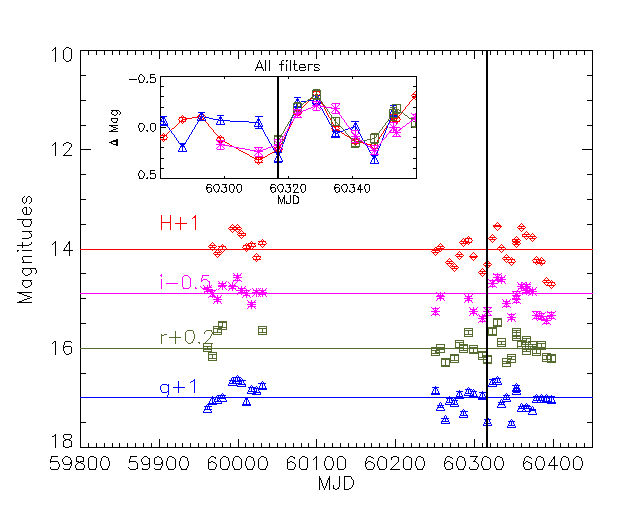} \\
   \caption{REM light curves of the blazar in the g, r, i and H filters. The photometric points were obtained after subtraction of the nearby star photometric template. The magnitudes were shifted for display purposes. Horizontal lines indicate the average magnitude values. The inset in the top of the figure shows the variation $\Delta$Mag = Mag-median(mag) for all filters around the time of the observation. In both plots, a vertical black line shows the date of our 2024 spectroscopic observation which happened near a local flux minimum.}
       \label{figREM}
  \end{figure}  
   In January 2023 we started monitoring the source using the 60~cm REM telescope \citep{Zer01} in the $g$, $r$, $i$, and $H$ filters. We 
   subtracted the magnitudes of our template from our measurements to obtain the blazar flux. It shows continuous variability correlated in all filters with roughly the same amplitude of $\pm0.15$ magnitudes (\autoref{figREM}). The average H magnitude before subtraction is $12.60\pm 0.03$, interestingly the 2MASS H magnitude is $H=12.34 \pm 0.03$ suggesting that the source was in a state slightly fainter than during 2MASS observations. 

  New FORS2 observations were performed on Jan 7, 2024, close to a local minimum of the blazar REM lightcurve (\autoref{figREM}).  The data reduction, flux calibration, and telluric corrections were performed as in our previous works \citep[see, e.g.][]{Gol21,Dam24}. Flux calibration is very difficult for this source due to the wavelength-dependent slit loss corrections and the contamination of the star, therefore we scaled the flux of the spectrum using our near-simultaneous REM photometry corrected with our photometric template.
  
  Spectra (see \autoref{figlines} and \autoref{figtotspec} in the Appendix) clearly show the presence of several emission features superimposed on the continuum and absorption features of the star. The weak, narrow feature of the 2022 observation is confirmed with much higher S/N while a weaker feature consistent with [OIII]a at $z\sim0.262$ appears at 6259~\AA. At the same redshift, the [NII]b, [NII]a and H$\alpha$ features (all narrow) are also detected. A broad, symmetric feature that can be identified with H$\alpha$ is clearly visible around 8300~\AA. All of these features are present before telluric corrections and in the single 2D and 1D frames of each observation. However, no emission consistent with H$\beta$, broad or narrow at or near this redshift, is detected.

    \begin{figure}[!]
   \centering
   \begin{center}$
\begin{array}{c}
   \includegraphics[scale=0.32]{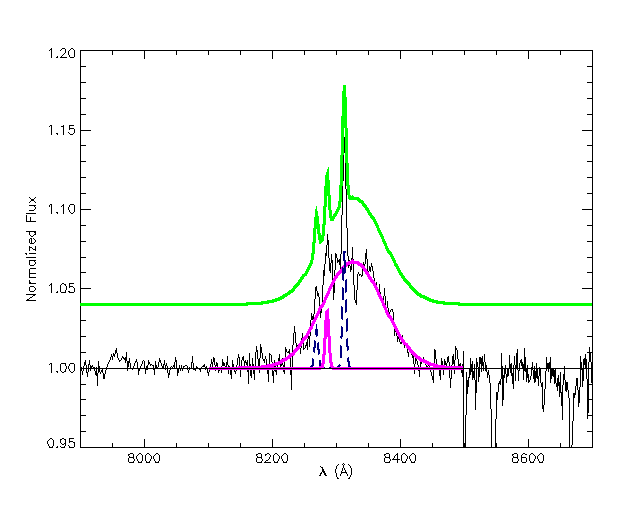}
   \end{array}$
      \end{center}
    \caption{H$\alpha$-[NII] complex doublet in the 2024 merged normalized FORS2 spectrum. The global four components fit is displayed in green, rescaled by 0.04 for visibility. The individual Gaussian components are also shown: in magenta the broad and narrow H$\alpha$ components and in dark blue (dashed lines) the [NII] components. While the three narrow components are at the systemic redshift $z=0.2621$, the broad and narrow H$\alpha$, both in magenta, have clearly different centroids. The absorption lines on the right of the plot are due to the Calcium Triplet of the nearby star.}
          \label{figlines}
   \end{figure}

    We fitted the [OIII] doublet using two Gaussians with independent widths and flux ratio. The H$\alpha$-[NII] complex was fitted  simultaneously with three narrow Gaussians features, constrained to have the same width, and a broad Gaussian feature. The narrow features are at a redshift consistent with the one of the [OIII] doublet but have a smaller width of $254 \pm 9$~km/s. The larger width of the [OIII] lines suggests the possible presence of an outflow as often is observed in these lines \citep[see e.g.][]{Sin22}. The signal-to-noise of the doublet is however too low (see the flux to error ratios in Table~\ref{tablines}) to investigate this possibility with a two-component fit (i.e. core + wing). We estimated the systemic redshift from the average position of the five narrow lines. Taking into account the uncertainties in the wavelength solution, a redshift value $z= 0.2621\pm0.0006$ is obtained which definitely disproves the value previously reported by \citet{Tho90}.
    The broad feature has EW$ =7.94\pm0.17$~\AA, greater than the limit of 5~\AA~usually quoted to distinguish between BL~Lacs and FSRQs \citep{Urr95}.
    The FWHM of the line is $4020 \pm 30$~km/s and its peak is shifted towards the red with respect to the narrow lines by $1460 \pm 36$~km/s. To verify this value, a simulated normalized spectrum of the target containing the H$\alpha$-[NII] complex was built. We then shifted only the broad H$\alpha$ by 30~km/s pixels between -2100~km/s and +2100~km/s and cross-correlated the resulting templates with the observed spectrum. The cross-correlation peak is found at $1500 \pm 30$~km/s, validating our result. The offset is thus detected at about the $40\sigma$ level and it is the most peculiar characteristic of this spectrum.

\begin{table}[!]
\small
\caption{\label{tablines} Properties of the emission lines detected.} 
\centering
\begin{tabular}{lccc}
\hline\hline
  Line     &   EW               &   Log(Lum)   &  FWHM  \\
              &~(\AA)                 &   (erg/s)          & (km/s)  \\ 
 \hline\hline             
  {[OIII]$\lambda$ 4959 } & $0.61\pm 0.04$  & $41.6\pm0.2$  & $693 \pm 50$  \\
  {[OIII]$\lambda$ 5007}  & $1.41\pm 0.10$  & $42.1\pm0.2$   & $520 \pm 13$  \\
 {[NII]$\lambda$ 6548}     & $0.18\pm 0.03$  & $40.9\pm0.2$ & $254 \pm 9$ \\
 {[NII]$\lambda$ 6583 }   & $0.53\pm 0.03$  & $41.6\pm0.2$ & $254 \pm 9$ \\
  {H$\alpha$ narrow}       & $0.27\pm 0.03$  & $41.3\pm0.2$ & $254 \pm 9$ \\
  {H$\alpha$ broad}         & $7.94 \pm 0.17$ & $43.0\pm0.2$  & $4020 \pm 30$  \\
\hline\hline
    
\end{tabular}

\end{table}

\section{Discussion}

The redshift we have found for PKS~0903-57, $z=0.2621$, is much lower than the commonly accepted value. It implies not only a lower distance but also much less $\gamma-\gamma$ absorption by the Extragalactic Background Light. We estimate a factor of $\sim1.25$ at 100~GeV and up to $\sim250$ at 1~TeV \citep{Dom11}. These appropriate corrections will greatly influence the inferred intrinsic VHE spectral properties of the object.

 The large EW of the broad line suggests that the source is not a BL~Lac. The existence of the broad Seyfert-like feature is also important for $\gamma$-ray spectral modeling as it proves the presence of an external photon field which is Comptonized by the electrons produced in the synchrotron jet \citep[see e.g.][]{Fin16}. 
 The Broad H$\alpha$ line luminosity is $1.0(\pm 0.5)$$\times 10^{43}$~erg/s. Assuming that it is powered by the thermal continuum of the AGN, this would imply a continuum luminosity $L_{5100} \sim 1.7\times10^{44}$~erg/s \citep[see eq. 1][]{Gre05}. However, we measure $L_{5100} \sim 4.1\times10^{45}$~erg/s in our spectrum which suggests that the continuum is dominated by another contribution, probably a jet.
  We thus can classify PKS~0903-57 as a FSRQ, the second nearest VHE FSRQ after the low-luminosity VHE FSRQ PKS~0736+017 \citep{HESS20}. We note however that, using {\it Fermi}-LAT spectra, in the plane Spectral Index -- Flux at 1~GeV PKS~0903-57 lies near the region occupied by Low Energy Peaked BL~Lacs \citep{Aje22} near the changing look blazar B21420+32 \citep{Mis21}.

  The most unusual feature of this spectrum is the offset towards the red of the broad H$\alpha$ feature from the systemic redshift by $1460 \pm 36$~km/s ($40 \pm 1$~\AA). This is the first time, to our knowledge, that such a symmetric feature is detected in a blazar in the optical range. We note that in the Black Hole (BH) Binary candidate OJ~287 a similar offset ($\sim 20$~\AA) was possibly detected \citep{Nil10} but the errors in the position of the line centroid prevented any claim on the robustness of this result.
  
 Doppler-shifted peaks may be produced by emission from an optically thick and geometrically thin relativistic Keplerian accretion disk; about 3--10\% of AGN display double-peaked broad lines \citep{Era94}. But Doppler boosting produces red and blue peaks that are usually asymmetric, only in extreme cases a single peak may be visible. In large sample studies, the majority of sources are found to have inclinations\footnote{$i=0^{\circ}$ for a face-on disk and $i=90^{\circ}$ for an edge-on disk} $15^{\circ}<i<30^{\circ}$ \citep{War24}. Given that for blazars the inclination angle is small ($\le 5$~degrees) and that the broad feature is symmetric, we consider below other possibilities.
  Several scenarios for this configuration can be proposed \citep[see e.g.][and references therein]{mar23}. One \citep{Wan17} assumes that the Broad Line Region is made of discrete clouds produced by clumps from the dusty torus. If the clumps are mostly falling into the Black Hole \citep[see Figure~1,][]{Wan17} a redshifted broad line may appear. The radial inflow would then face a strong radiation field pushing away the infalling gas. The gas should likely become turbulent and the line profile become irregularly variable over time spans shorter than the ones detected with reverberation mapping \citep[few days, see e.g.][]{Pet97}. 
  Alternatively, two Supermassive Black Holes might be in the process of merging with the Broad-line emission generated around one BH and the narrow lines generated around the other \citep{Dot09}. Many kinematically offset binary BH candidates have been proposed \citep[see e.g.][and references therein]{Kel21}. In this case, orbital motions are very difficult to detect except for very small separations ($< 1$~pc). Therefore the configuration is stable, for example for observations spaced by five years or less, no velocity variation is expected to be detectable \citep{Kel21}.
  
  Finally, the most exciting possibility is that we are observing a recoiling Black Hole formed after the coalescence of two BHs. The emission of gravitational waves during the merging is anisotropic and therefore the resulting Black Hole may receive a kick with velocities up to a few thousands of km/s \citep{Cam07} and be displaced from the center of the host galaxy or even ejected altogether. One of the best candidates for this scenario is 3C186 \citep{Chi18}. Hubble Space Telescope Imaging Spectrograph (HST/STIS) and Sloan Digital Sky Survey spectroscopy discovered compatible blueshifts ($\sim 2000$~km/s) from the systemic redshift in several emission lines while HST Wide Field Camera imaging revealed a 1\farcs3 offset ($\sim11$~kpc at the redshift of the source, $z=1.06$). In this case, deep imaging is key to detecting and separating host and AGN emission.

\section{Conclusions}

 The nature of PKS~0903-57 has been hidden for a long time by the presence of a bright ($G\sim 16$) star at just 0\farcs67. Taking advantage of appropriate slit positioning and exceptional seeing, we were able to detect the blazar emission, our main results are:
 
$\bullet$ We provide the first solid measurement of the redshift, lowering its distance estimate by a factor of three, with strong implications for understanding the source emission.

$\bullet$ A conspicuous broad line was detected along with a very bright continuum suggesting that PKS 0903-57 is a very rare low-luminosity, low-redshift VHE FSRQ.

$\bullet$ We found an unprecedented (for blazars) velocity offset between the systemic redshift and the broad line which suggests either a very peculiar accretion state or that we are witnessing the consequence of a merging.

 In order to better understand the nature of this very intriguing system, further observations are needed. On the one hand, longer spectroscopic observations in exceptional seeing will make it possible to detect H$\alpha$ again with higher signal-to-noise and to check for variability in its profile. Such spectra will also possibly detect broad H$\beta$ and compare its profile to the one of H$\alpha$, giving further information on the Broad Line Region properties. On the other hand, deep high-resolution imaging observations will enable, for the first time, to separate the active nucleus and the star.  Given the relatively low redshift, a detection of the host galaxy and of the putative merging system is also possible.

\begin{acknowledgements}
We thank the observers for performing our observations at SALT and VLT, in particular, John Pritchard at ESO for very productive discussions. This research has made use of the SIMBAD database, operated at CDS, Strasbourg, France.This paper went through internal review by the CTAO consortium. We thank Elina Lindfors and Nicola Masetti for their helpful comments and suggestions as internal reviewers. This research has made use of the CTAO instrument response functions, see \url{https://www.cta-observatory.org/science/cta-performance} (version prod3b-v1) for more details. We gratefully acknowledge financial support from the agencies and organizations listed here: \url{http://www.cta-observatory.org/consortium_acknowledgments}, and in particular the U.S. National Science Foundation Grant PHY-2011420 and ANID/Chile FB210003.
\end{acknowledgements}

\bibliographystyle{aa}
\bibliography{pks} 

\begin{appendix} 
\onecolumn
\section{Field of view image and complete spectra}

 \begin{figure*}[!h]
     \centering
  \includegraphics[scale=0.4]{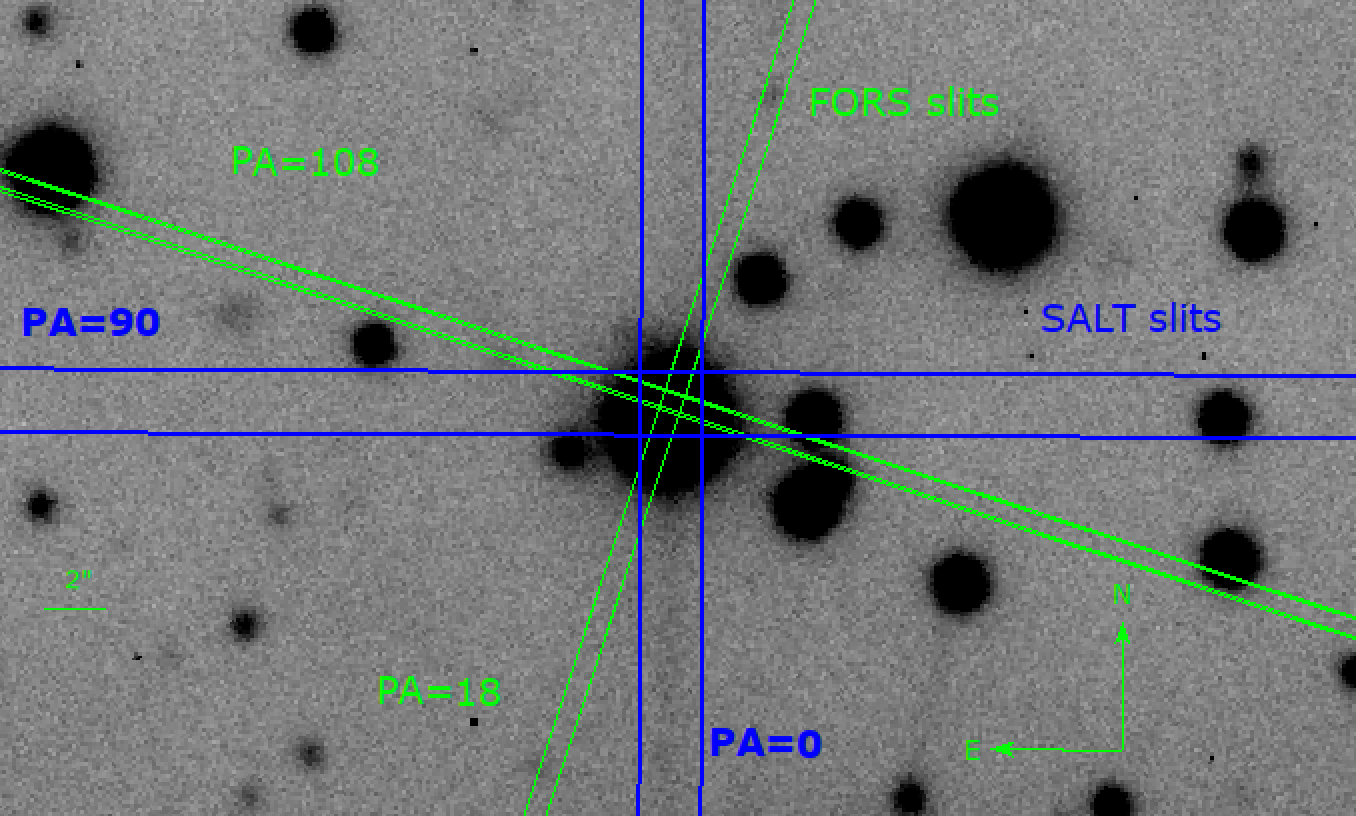}
 \caption{Slit positions of our observations superimposed on a SUperb Seeing Imager R-band image of the field (observation date 1994-01-08) retrieved from the ESO archive. Note that, as the star is much brighter than the blazar in this observation, the centers of our observations are offset from the center of the optical emission. The blue lines represent the SALT slits, (width=2\arcsec) and the green lines represent the FORS2 slits (width=0.7\arcsec and 0.5\arcsec). Position angles (in degrees) are also indicated near the corresponding slits with corresponding colors. }
           \label{figslits}
   \end{figure*}

\begin{figure*}[!h]
     \centering
     \includegraphics[scale=0.4]{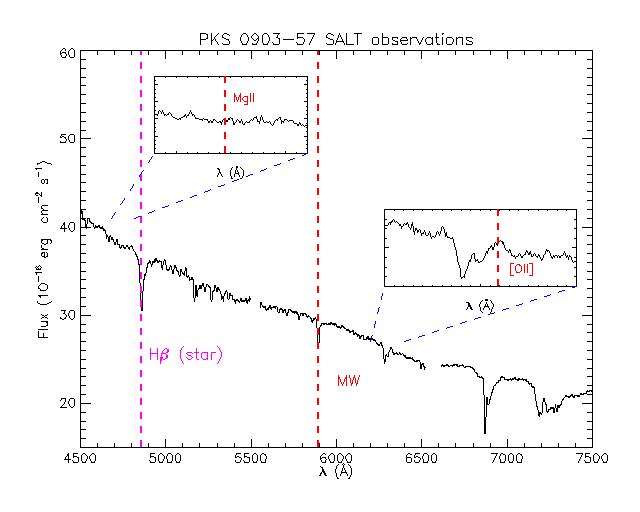}
 \caption{SALT spectrum of our 2020 observation of PKS 0903-57. The spectrum shows prominent stellar absorption lines and a strong NaID interstellar absorption feature. No emission feature possibly linked to the AGN is detected. In particular, we show in two insets the zoom of the wavelength regions where the features detected by \citet{Tho90} should lie. Only weak ripples are visible.
 }
          \label{figSALT}
   \end{figure*}    

  \begin{figure*}[!h]
     \centering
    \includegraphics[scale=0.4]{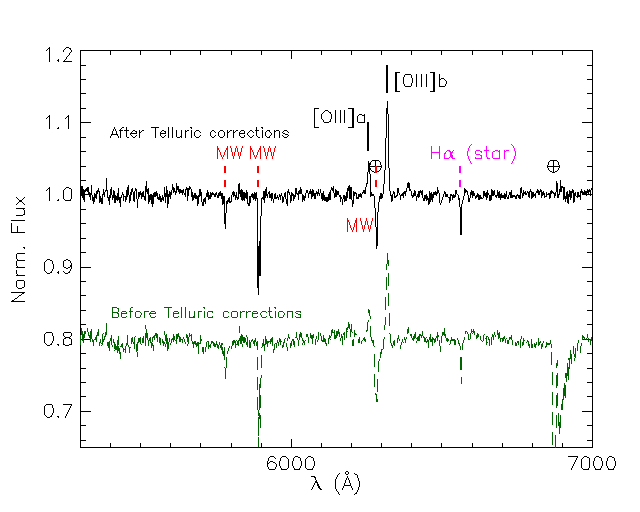}
    \includegraphics[scale=0.4]{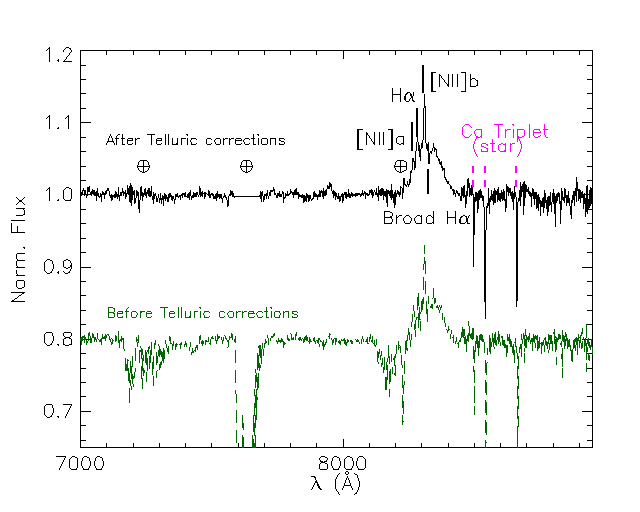}
    \caption{Normalized and telluric corrected spectrum of our 2024 FORS2 observations of PKS 0903-57 (continuous black line) and normalized spectrum of our 2024 FORS observations of PKS 0903-57 before telluric corrections (dashed green line). \emph{Left panel:} spectra between 5300 and 7000~\AA; \emph{Right panel:} spectra between 7000 and 8950~\AA.
    The AGN features ($z=0.2621$) are marked with continuous lines (black), the stellar features ($z=0$) are marked with dashed lines (magenta). Telluric absorptions are marked by the symbol $\oplus$ and Galactic absorption features are labelled with `MW' and a dashed line (red).  A possible feature (EW $\sim 0.4$~\AA), visible around 7950~\AA~would be consistent with [OI] $\lambda$ 6300 at $z=0.2621$.
    }
          \label{figtotspec}
   \end{figure*}

\end{appendix} 

\end{document}